# UAV and Machine Learning Based Refinement of a Satellite-Driven Vegetation Index for Precision Agriculture [†]

**Vittorio Mazzia** [1,2]**, Lorenzo Comba** [3,4]**, Aleem Khaliq** [1,2,*]**, Marcello Chiaberge** [1,2] **and Paolo Gay** [3]

[1] Department of Electronics and Telecommunications, Politecnico di Torino, Corso Duca degli Abruzzi 24, 10129 Torino, Italy; vittorio.mazzia@polito.it (V.M.); marcello.chiaberge@polito.it (M.C.)
[2] PIC4SeR, Politecnico Interdepartmental Centre for Service Robotics, 10129 Turin, Italy
[3] Department of Agricultural, Forest and Food Sciences, Università degli Studi di Torino, Largo Paolo Braccini 2, 10095 Grugliasco (TO), Italy; lorenzo.comba@unito.it (L.C.); paolo.gay@unito.it (P.G.)
[4] Institute of Electronics, Computer and Telecommunication Engineering of the National Research Council of Italy, c/o Politecnico di Torino, Corso Duca degli Abruzzi 24, 10129 Torino, Italy
* Correspondence: aleem.khaliq@polito.it; Tel.: +39-011-0903551
† This paper is an extended version of the conference paper: Khaliq, A.; Mazzia, V.; Chiaberge, M. Refining satellite imagery by using UAV imagery for vineyard environment: A CNN Based approach. In Proceedings of the IEEE International Workshop on Metrology for Agriculture and Forestry (MetroAgriFor), Portici, Italy, 24–26 October 2019; pp. 25–29.

**Abstract:** Precision agriculture is considered to be a fundamental approach in pursuing a low-input, high-efficiency, and sustainable kind of agriculture when performing site-specific management practices. To achieve this objective, a reliable and updated description of the local status of crops is required. Remote sensing, and in particular satellite-based imagery, proved to be a valuable tool in crop mapping, monitoring, and diseases assessment. However, freely available satellite imagery with low or moderate resolutions showed some limits in specific agricultural applications, e.g., where crops are grown by rows. Indeed, in this framework, the satellite's output could be biased by intra-row covering, giving inaccurate information about crop status. This paper presents a novel satellite imagery refinement framework, based on a deep learning technique which exploits information properly derived from high resolution images acquired by unmanned aerial vehicle (UAV) airborne multispectral sensors. To train the convolutional neural network, only a single UAV-driven dataset is required, making the proposed approach simple and cost-effective. A vineyard in Serralunga d'Alba (Northern Italy) was chosen as a case study for validation purposes. Refined satellite-driven normalized difference vegetation index (NDVI) maps, acquired in four different periods during the vine growing season, were shown to better describe crop status with respect to raw datasets by correlation analysis and ANOVA. In addition, using a K-means based classifier, 3-class vineyard vigor maps were profitably derived from the NDVI maps, which are a valuable tool for growers.

## 1. Introduction

Precision agriculture is considered to be a fundamental approach to pursue a low-input, high-efficiency, and sustainable agriculture [1,2] which implements new technological solutions [3,4]. For precision agriculture to be effective, however, a reliable description of the local status of the crops is essential to perform site-specific management practices when using automatic machinery and even robotics [5–7]. To this extend, the relevance of remote sensing has widely been demonstrated for the extension of in-field surveys to entire plots or even regions [8–11]. This is particularly true for satellite imagery, which has profitably been exploited for in-field mapping [12,13], crops status monitoring [14,15], and disease assessment [16], both spatially and temporally [17].

However, freely available satellite imagery with low or moderate resolution showed some limits in specific applications, resulting in it being not directly suitable for field monitoring purposes in some agricultural contexts [18,19], such as orchards and vineyards. Indeed, detailed crop information is usually required in these contexts [20], provided by computing crop status indexes, such as the normalized difference vegetation index (NDVI) [21], even at the plant scale [22]. The presence of different elements in these scenarios, such as crops and terrain (inter-row space, in the case of crops grown in rows), causes pixels with mixed natures in low resolution satellite imagery, which can lead to biased crop indices [18].

A profitable approach to improve the performance of remote sensing by satellite data is the exploitation (and fusion) of information from additional data sources, such as agrometeorological data [23], in situ plot data [24], laser altimetry data [25], thermal imagery [26], or even the concurrent use of different satellite platforms [27]. Zhao et al. proposed the fusion of data acquired from Unmanned Aerial Vehicle (UAV) and satellite based sensors to improve crop classification [28]. Many efforts have also been made to increase the quality of moderate resolution platforms with advanced computing techniques, such as the super-resolution approach based on machine learning, with deep neural networks (DNN) and convolutional neural networks (CNN) being the most exploited ones [29–32]. For example, several convolutional network architectures were proposed to enhance the spatial details of drone-derived images [33]. Indeed, an intrinsic capability of deep learning is distributed learning, which distributes, among all the variables of the model, the knowledge of the dataset and the capability to extract such high-level, abstract features [34]. Altogether, it provides deep learning with the ability to learn more robust mapping functions with much more generalization power than traditional machine learning algorithms [35]. In addition, data augmentation techniques further increase their performance [36].

With this approach, new methods aimed at synergically exploiting freely available satellite imagery, refined by high-resolution UAV-based datasets, can be highly effective [37,38]. Few studies have been performed on satellite imagery improvements based on centimetric imagery acquired from UAVs, such as the estimation of canopy structures and biochemical parameters [39] and the estimation of macro-algal coverage in the yellow sea by refining satellite imagery using high resolution airborne based synthetic aperture radar (SAR) imagery [40]. The problem of the spatial dynamics of invasive alien plants was profitably solved by [41], merging single- and multi-date UAV and satellite imagery. In [42], a UAV-based inversion model was applied to the satellite's imagery with reflectance normalization to monitor the salinity in coastal saline soil.

However, new approaches should be conceived to refine low resolution satellite imagery, which should be freely available and with a short revisiting time, by means of the mapped spatial information of high-resolution imagery from sporadic, or even single, UAV flights. This approach could improve the reliability of remotely sensed satellite data in complex scenarios, such as vineyards, making it highly cost-effective.

In this work, a novel approach to refine moderate resolution satellite imagery by exploiting information properly derived from UAV-driven high-resolution multispectral images is presented. The proposed method, based on deep learning techniques, is able to provide enhanced decametric NDVI maps of vineyards from frequent and freely available moderate resolution satellite imagery. To train the convolutional neural network, only a single UAV-driven dataset is required, making the proposed approach simple and cost-effective. In addition, by using a K-means-based classifier, 3-class vineyard vigor maps were profitably derived from the NDVI maps, which are a valuable tool for growers. For validation purposes, a vineyard in Serralunga d'Alba (Northern Italy) was chosen to perform this study, which involved three parcels and four different time periods, during the whole vine growing season. Refined satellite-based NDVI maps were shown to better describe crop status with respect to the raw datasets. The manuscript is organized as follows: the proposed satellite NDVI refinement method is presented in Section 2, together with its architecture and supervised training phase; Section 3 presents the experimental case study, the performed validation approach and the obtained results; and finally, Section 4 reports the conclusions.

**2. Methods**

The refinement framework developed in this study is aimed at increasing the reliability of the decametric NDVI maps of vineyards derived from freely available satellite imagery. It is based on a convolutional-based neural network (CNN) architecture, hereafter called RarefyNet, which is capable of learning feature representations with a supervised approach, after a training phase. The RarefyNet, taking advantage of compositionality, is able to extract in a hierarchical manner features from its input data and exploit its internal knowledge to obtain a refined value of its input samples. To train the RarefyNet, a single UAV-driven dataset was used as reference. Indeed, NDVI maps from UAV airborne sensors were shown to be more reliable than raw moderate resolution satellites in describing actual crop status [18]. Once trained, the RarefyNet can refine the satellite-driven decametric NDVI maps of the vineyard acquired in any time period during the vine growing season. In addition, using a K-means based classifier, vineyard maps with three vigor classes (low, medium, and high vigor) were profitably derived from the NDVI maps, which are a valuable tool for growers. The mathematical notation adopted in the following is summarized in Table 1.

**Table 1.** Adopted mathematical notation.

| Variable | Definition |
|---|---|
| $a$ | a vector |
| $A$ | a matrix |
| $\mathbf{A}$ | a tensor |
| $a_i$ | i-th element of a vector $a$ |
| $A_{i,j,k}$ | element i,j,k of a 3-D tensor $\mathbf{A}$ |
| $A_{:,:,i}$ | 2-D slice of a 3-D tensor $\mathbf{A}$ |
| $X$ | a set of elements/a map |
| $X^{(i)}$ | i-th sample from a dataset |
| $y^{(i)}$ | ground-truth associated with the i-th sample |

*2.1. RarefyNet: Input, Output, and Architecture*

Considering a decametric NDVI map $X_{raw}$ from a raw satellite dataset, constituted by pixels $x_i \in X_{raw}$, the pixels $\hat{y}_i$ of an enhanced NDVI map $\hat{X}$ can be generated by the RarefyNet's non-linear mapping function with parameters $\Theta$ as:

$$\hat{y}_i = F(X^{(i)}, \Theta) \qquad (1)$$

where $X^{(i)}$ is an input tensor derived from $X_{raw}$. Input tensor $X^{(i)}$ was defined to collect information, in terms of the NDVI digital value and position on the map, on pixel $x_i$ and on a subset of its neighbourhood. Indeed, the contribution of a map pixel is strictly related to its relative position with respect to its surrounding pixels. In more detail, input tensor $X^{(i)}$ was thus defined as a three-dimensional tensor $X \in R^{3 \times 3 \times 2}$, where the first layer is a 3 × 3 map patch (formally $X^{(i)}_{:,:,0}$), centered in $x_i$ (formally element $X^{(i)}_{1,1,0} = x_i$), and the second layer ($X^{(i)}_{:,:,1}$) is made of the set of unique location values of map pixels $X^{(i)}_{:,:,0}$ in the first layer, defined as the linear indexing of the raster matrix. Of course, in order to also consider boundary pixels, a zero-padding operation was performed on the overall maps to allow tensor extraction in boundary pixels. That does not influence the behavior of later feature maps of the network.

A graphical representation of the overall RarefyNet architecture is illustrated in Figure 1. Inspired by [43,44], input tensor $X^{(i)}$ feeds a stack of two inception blocks that gradually extract the spatial correlation between the 8 neighborhood pixels and central target pixel $X^{(i)}_{1,1,0}$. The features of NDVI map $X_{:,:,0}$ are concurrently processed by an ensemble of parallel convolutional layers with the same number of filters $n$, but different filter sizes $f$ and dilatation rates $k$. Indeed, distinct kernel sizes extract different correlations from the data and, on the other hand, Atrous convolutions take advantage of non-local spatial correlations. Finally, batch normalization [45], as a regularization technique, is applied to each branch before an exponential linear unit (ELU) [46] activation and final concatenation along the feature dimension. Zero padding is applied before each module in order to preserve the first two dimensions of the input tensor. Starting with the first inception block, an input patch $X^{(i)}$ with shape $(3, 3, 2)$ is concurrently processed by the ensemble of parallel convolutions producing an output tensor of shape $(3, 3, n_I)$ where $n_I$ is the result of the feature map concatenation of the different convolutional branches. The second inception module builds on top of this feature tensor by constructing further high-level representations and generating a multi-dimensional array with $n_{II}$ feature maps.

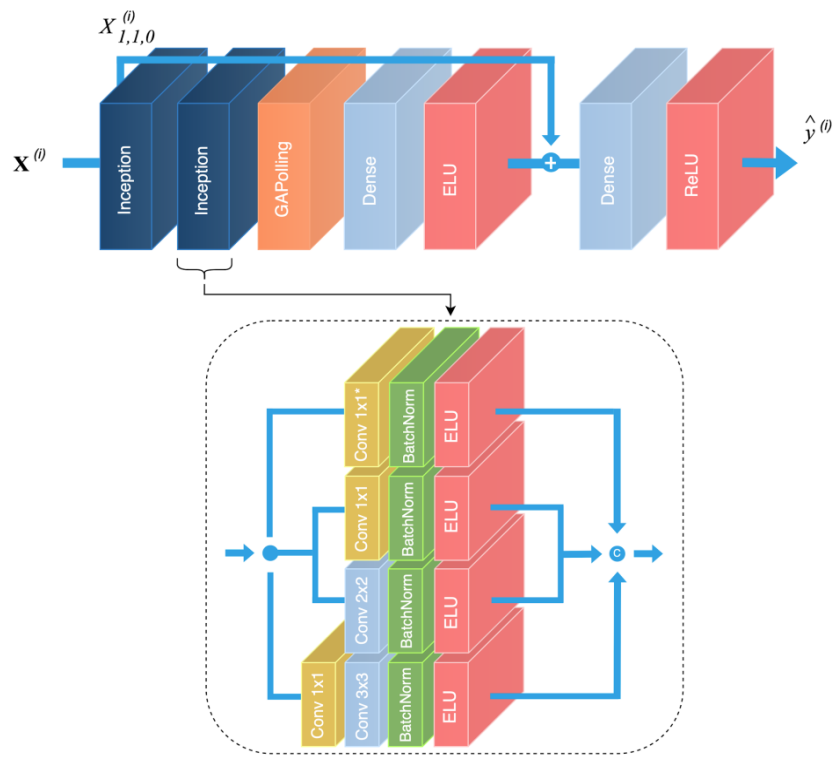

**Figure 1.** Graphical representation of the proposed RarefyNet model. The overall residual architecture is depicted in the top part of the figure with a detailed overview of its inception modules. Input tensors are processed by two inception modules that build their representations on top of each other, concatenating outputs of their different branches.

The output tensor produced by the cascade of inception blocks feeds a global average pooling (GAP) layer which reduces the rank of the input tensor producing a 1-D output array. The GAP operation reduces the spatial dimension of its input tensors, reinforcing the feature maps to be confidence maps of concepts. The GAP 1-D output array feeds a fully connected layer that terminates with a single unit with the ELU as an activation function. The ELU brings non linearity to the model, but still produces both positive and negative values. At this stage, a residual connection sums the output of the dense layer with the original NDVI pixel $X^{(i)}_{1,1,0}$ to be refined. The residual connection, inspired by super-resolution neural network architectures, covers a primary role inside the overall model; it largely simplifies the role of the first part of the network by moving its objective towards a mere refining operation of the satellite's input pixel. Indeed, the model does not have to recreate the value of the input pixel after processing of the convolutional filters, but progressively learns from ground truths how to use the starting satellite input value with its eight neighbors to estimate the inter-row radiometric contributions and refine the raw decametric NDVI value $x_i$. Finally, a second fully connected layer with rectified linear units (ReLU) with activation functions produces output prediction $\hat{y}^{(i)}$ by removing any off-set between the satellite and the UAV NDVI spaces.

The complementary use of (1) a deep learning-based architecture, of (2) different regularization techniques to constrain the space parameter, and of (3) a 1 × 1 convolution to reduce the number of model parameters, produces a light-weight and efficient solution to construct a complex non-linear map between satellite and enhanced UAV pixel information.

*2.2. RarefyNet: Training Phase*

To identify an effective set of parameters $\Theta$, the RarefyNet model (Equation (1)) has to be trained. The training phase is an iterative process during which parameters $\Theta$ are adjusted to reduce the error defined as the difference between the desired refined NDVI values $\hat{y}$ and reference value $y$. In this application, the enhanced NDVI map $Y_{UAV} = \{y_i\}$ derived from the UAV flights was adopted as the reference dataset for the training phase. In particular, the UAV-driven $Y_{UAV}$ dataset was derived by detecting vineyard canopies within the high resolution imagery and by a proper down-sampling procedure, described in detail in [18]. The defined training samples are thus made by the properly paired tensors $X^{(i)}$, from raw satellite-driven NDVI pixel $x_i$, and a reference NDVI $y_i$, from the accurate UAV-driven dataset. Moreover, in order to enlarge the number of available training examples and consequently reducing possible overfitting problems, a simple data augmentation technique was applied; considering the ith sample and maintaining the central satellite pixel $X_{i,1,1,:}$ fixed, it is possible to produce $(K-2)$ new samples from each original training data point by rotating the other eight pixels around the central one.

During the training phase, a loss function $\mathcal{L}$ based on the norm-2 measure

$$\mathcal{L} = \left(\frac{1}{m}\left(\sum_{i=1}^{m}|\hat{y}^{(i)} - y^{(i)}|^2\right)\right)^{\frac{1}{2}} \quad (2)$$

of the difference between model output predictions $\hat{y}^{(i)}$ and reference $y^{(i)}$ will be used together with a mini-batch gradient descent method and $m$ training instances to optimally identify the parameters $\Theta$ of the network. The loss function $\mathcal{L}$ is a typical performance measure for regression problems and it estimates how much error the model typically makes in its predictions, with a higher weight for large errors. Model training is therefore performed iteratively by feeding the network with a batch of a certain dataset size and updating the parameters with small steps which are determined by learning rate $\eta$, by using the gradient of the selected loss function.

*2.3. RarefyNet: Structure Optimization*

The final architecture, shown in Figure 1, is thus the result of a careful design aimed at obtaining the best performance in terms of reliability and computational costs. The final model is a light-weight neural network architecture with 16,296 trainable parameters.

Every inception block has four parallel branches with different filter sizes $f$ and dilatation rates $k$. In the first branch (bottom of Figure 1), the 1 × 1 convolution halves the number of feature maps in order to reduce the number of parameters and the computational requirements by the following convolutional layer. The first inception module produces eight feature maps for each branch, which are linked in a unique output tensor with $n_I$ channels after being separately pre-processed by a batch normalization layer and an ELU activation function. Equally, the second inception block produces $n_{II} = 32$ feature maps for each branch, which are linked in a final tensor that feeds the GAP layer. Subsequently, a fully connected layer reduces the 1-D output tensor first to 32 and then to 1 before feeding the residual connection. Moreover, a dropout layer, with $p = 0.2$, is inserted between the two fully connected layers in order to regularize the network and produce a very robust and reliable model [47]. Finally, an output neuron, with an ReLU activation function, closes the head of the network in order to compensate and mitigate the presence of possible biases.

The technique proposed by Smith et al. in [48] was adopted to identify the maximum value of learning rate $\eta = 5 \times 10^{-4}$ to start with. Finally, beside batch normalization and dropout, the AdamW [49] updating rule

$$\Theta_{t+1} = \Theta_t - \frac{\eta}{\sqrt{\hat{v}_t + \varepsilon}} m_t - \eta \alpha \Theta_t \tag{3}$$

was used, which is a modified version of the well-known Adam optimizer [50] with L2 regularization, where $m_t$ and $\hat{v}_t$ are the exponential decay of the gradient and gradient squared, respectively, and $\alpha$ is a new regularization hyperparameter to be set for the learning process. This is a simple fix to the classic updating rule of the Adam optimizer, but it has repeatedly shown far better results than the L2 regularization for all experimentations.

In order to find the best training hyperparameters for the optimizer and the network, we used 10% of the training set to perform a random search evaluation, with few optimization iterations, in order to select the most promising parameters. Then, after this first preliminary phase, the analysis focused only on the most promising hyperparameter values, fine tuning them with a grid search strategy.

*2.4. Vigor Classifier*

Using an unsupervised clustering algorithm, satellite pixels $x_i \in X_{raw}$, RarefyNet predictions $\hat{x}_i \in \hat{X}$ and down-sampled UAV pixels $y_i \in Y_{UAV}$ were classified into three different vigor classes: low, medium, and high. In particular, a K-means clustering algorithm was separately fitted on the three NDVI maps by using Elkan's algorithm and k-means++ to initialize the centroids. Each fitting was run 15 consecutive times with a maximum of 500 iterations and a tolerance of $10^{-4}$. The outputs with the lowest within cluster sum of squared (WCSS) distance were selected as the final clustered maps of the three NDVI sets.

**3. Experiments and Results**

The effectiveness of the proposed approach to refine moderate resolution imagery by using UAV-driven imagery was tested in the vineyard selected as the case study. The RarefyNet was implemented in the TensorFlow framework [47,51] and trained with satellite and UAV-based datasets acquired in May 2017 (time I). For validation purposes, the trained RarefyNet was used to enhance the NDVI map from the satellite platform acquired in three different time periods (June, July, and September: time II, III, and IV) and the results were compared with the more accurate UAV-driven NDVI maps.

In more detail, the study was conducted in a vineyard located in Serralunga d'Alba, Piedmont, in the northwest of Italy, shown in Figure 2. The selected area includes three parcels, with a total surface of about 2.5 hectares. The area is located at approximately 44°62′4″ latitude and 7°99′9″ longitude in the World Geodetic System 1984. The test site elevation is within the range of 330 to 420 m above sea level, with steep slope areas (about 20%). Parcels are cultivated with the cultivar Nebbiolo grapevine. The vineyard soil is predominantly loamy. The irregularity of the terrain's morphology, in terms of altitude, slope, and soil exposure to the sun, affects microclimatic conditions and water availability within and between parcels [20].

*3.1. Satellite and UAV-Based Time Series Imagery*

In this study, cloud-free level-2A Sentinel-2 bottom of atmosphere (BOA) reflectance images were used as moderate resolution satellite imagery. Sentinel-2 data products were downloaded from the Copernicus open access hub and imported into a processing platform SNAP toolbox (6.0) provided by European Space Agency (ESA). By using subset command in SNAP, pixels of the Sentinel-2 images were extracted in accordance with the study cite. Geometric, atmospheric, and Bidirectional Reflectance Distribution Function (BDRF) corrections were performed by using a Sen2cor processor, which is a plugin for SNAP [52–54]. More details about Sentinel-2 products can be found in [55]. The selected satellite tiles were acquired on four dates during the 2017 growing season (Table 2) in order to consider different vegetative vine statuses. Only red and near infrared bands (bands 4 and 8, respectively), that match with the spectral channels of UAV airborne sensors, were used in this study (with ranges 650–680 nm and 785–900 nm, respectively) to produce the NDVI maps [4,5,8], widely used for vegetation monitoring and health assessment of crops. The pixels that were completely included within the boundaries of the three considered "Parcel A", "Parcel B" and "Parcel C" were selected, as shown in Figure 2a.

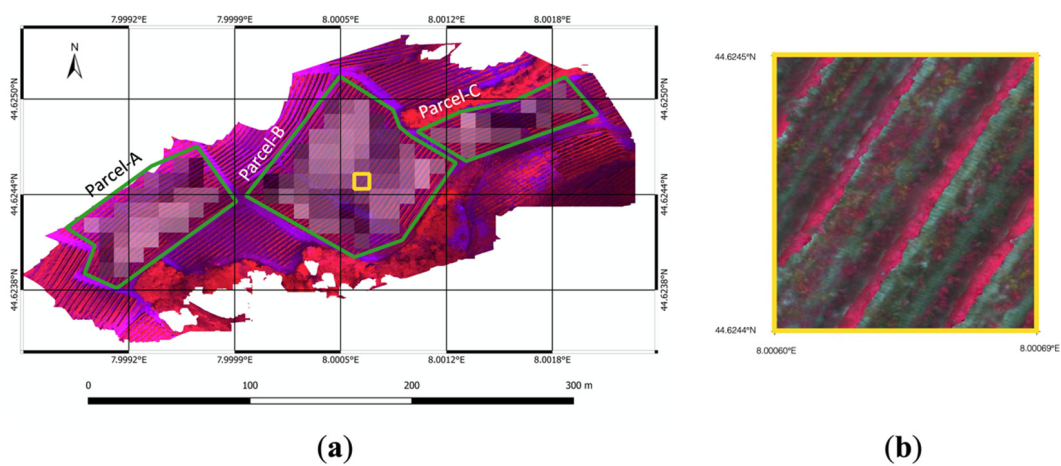

**Figure 2.** (**a**) Selected test field located in Serralunga d'Alba (Piedmont, northwest of Italy). The boundaries of the three considered parcels, named "Parcel-A", "Parcel-B", and "Parcel-C", are marked with solid green polygons. The concurrent illustration of low resolution and high-resolution maps derived from satellite and UAV respectively is represented in false colors (near infrared, red, and green channels). (**b**) Enlargement of UAV imagery highlighted by the yellow square in Figure 1a.

**Table 2.** Dataset acquisition details from the Sentinel-2 ($X_{raw}$) and UAV ($Y_{UAV}$) platforms.

| Time Period | Dataset Name | Acquisition Date | Source |
|---|---|---|---|
| I | $X_{raw}^{I}$ | 30 April 2017 | Sentinel-2 |
| | $Y_{UAV}^{I}$ | 5 May 2017 | UAV |
| II | $X_{raw}^{II}$ | 6 July 2017 | Sentinel-2 |
| | $Y_{UAV}^{II}$ | 29 June 2017 | UAV |
| III | $X_{raw}^{III}$ | 5 August 2017 | Sentinel-2 |
| | $Y_{UAV}^{III}$ | 1 August 2017 | UAV |
| IV | $X_{raw}^{IV}$ | 17 September 2017 | Sentinel-2 |
| | $Y_{UAV}^{IV}$ | 13 September 2017 | UAV |

The decametric UAV-based NDVI maps, used as accurate references, were derived from red and near infrared bands (with ranges 640–680 nm and 770–810 nm, respectively) of high-resolution multispectral imagery acquired by a UAV airborne Parrot Sequoia® multispectral camera. The UAV path was planned to maintain flight height close to 35 m with respect to the terrain by properly defining waypoint sets for each mission block on the drone guidance platform based on the GIS cropland map. With this specification, the aerial images ground sample distance (GSD) turned out to be 5 cm (Figure 2b). The UAV flights were performed on four different dates over the 2017 crop season (Table 2), according to the satellite's visiting dates. The high-resolution multispectral imagery was then processed to select only the pixels representing vine canopies and was down-sampled to be in accordance with the satellite's spatial resolution (as described in [18]), obtaining UAV-driven decametric NDVI map $Y_{UAV}$.

*3.2. Experimental Settings*

The RarefyNet used in this experimentation was trained with training tensors derived from raw dataset $X_{raw}^{I}$ and decametric NDVI map $Y_{UAV}^{I}$, which were acquired in May (time I). In more detail, after the sample extraction procedure and the data augmentation process were applied to the training samples (Section 2.1), a set of 1379 and 591 tensors were obtained for the training and test procedures, respectively. The proposed architecture was trained for 300 epochs with a batch size of 64. No learning rate strategies were applied, but the value of the learning rate was kept constant for all the training epochs of the optimization procedure. All tests were carried out with the TensorFlow framework on a workstation with 64 GB of RAM, an Intel Core i7-9700K CPU and an Nvidia 2080 Ti GPU.

Since, at the agronomical scale, maps of classes with different vigor levels can be derived by an expert in-field survey, the validation of the NDVI map refinement was performed by assessing their conformity to a three-level vigor map. Thus, a preliminary validation was performed by feeding the trained RarefyNet model with satellite-driven raw map $X_{raw}^{II}$ (time II) and the obtained output, in the form of refined map $\hat{X}^{II}$, was compared with reference map $V_{field}^{II}$ produced by the in-field survey [18]. For completeness, the effectiveness of satellite-driven raw map $X_{raw}^{II}$ and UAV-driven NDVI map $Y_{UAV}^{II}$ in discriminating vigor levels described in $V_{field}^{II}$ was also investigated.

To extend validation to other time periods (time I, III and IV), three-level vigor maps $Y_{UAV}$ were derived by applying the K-means algorithm to UAV-driven dataset $Y_{UAV}$, to be used as the ground truth reference. Indeed, the soundness of this approach was confirmed by validating the selected classifier with the dataset of time II, clustering $Y_{UAV}^{II}$, and comparing it with ground truth vigor map $V_{field}^{II}$ (Figure 3).

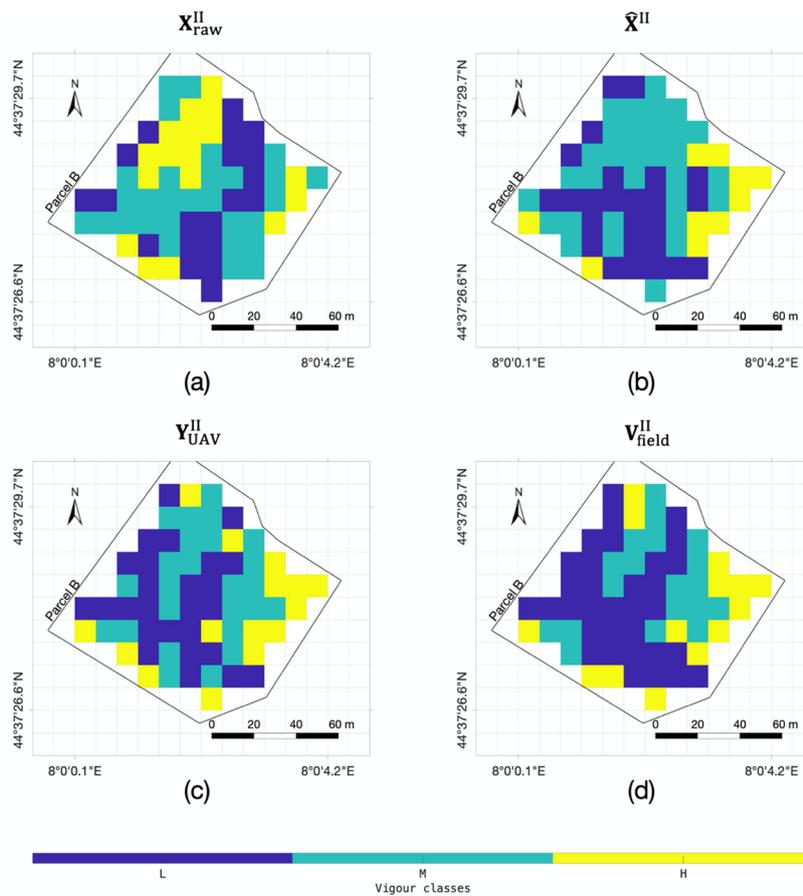

**Figure 3.** Three-level vigor maps (**a**) $X_{raw}^{II}$, (**b**) $\hat{X}^{II}$, and (**c**) $Y_{UAV}^{II}$ of parcel B, derived from raw Sentinel-2 NDVI map $X_{raw}^{II}$, refined satellite NDVI map $\hat{X}^{II}$ and UAV-driven NDVI map $Y_{UAV}^{II}$, respectively. Vigor map (**d**) of parcel B from the expert's in-field survey $V_{field}^{II}$. Maps $X_{raw}^{II}$, $\hat{X}^{II}$ and $Y_{UAV}^{II}$ were obtained by the selected K-means based classifier.

With this approach, the validation of the temporal effectiveness of the proposed satellite-driven dataset refinement framework was performed by refining datasets $X_{raw}^{I}$, $X_{raw}^{III}$, and $X_{raw}^{IV}$ and assessing the accordance between the obtained refined NDVI maps ($\hat{X}^{I}$, $\hat{X}^{III}$, and $\hat{X}^{IV}$) and the UAV-driven reference ones ($Y_{UAV}^{I}$, $Y_{UAV}^{III}$, $Y_{UAV}^{IV}$).

*3.3. Results and Discussion*

NDVI maps derived from onboard UAV sensors are used in many agricultural applications due to their effectiveness in providing high spatial resolution imagery and control over the data acquisitions [20–22]. However, there are constraints such as limited flight time of UAVs, labor extensiveness, and lower coverage that make it less affordable than satellite imagery. In contrast, NDVI maps derived from the satellite-based sensors have been widely used for the past four decades [56]. The latest developments in the satellite-based sensors provide frequent imagery with fine spectral information and moderate spatial details. However, satellite based remote sensing for vegetation monitoring becomes more challenging when considering crops with discontinuous layouts, such as vineyards and orchards [57]. The primary reason behind this is the presence of inter-row paths and weed vegetation within the cropland, which may deeply affect the overall spectral indices computation, leading to a biased crop status assessment. Therefore, refinement of the satellite driven vegetation index is performed in this study.

The effectiveness of the refined NDVI map $\hat{X}^{II}$, generated by the trained RarefyNet model, in describing the actual vigor status of the vineyard selected as the case study was investigated by performing ANOVA between map pixels properly grouped based on the vigor classes expressed in $V_{field}^{II}$, selected as the ground truth (Figure 3d). In order to demonstrate the obtained improvement, the coherence of raw satellite-driven map $X_{raw}^{II}$ and of UAV-driven NDVI map $Y_{UAV}^{II}$ with the ground truth was performed. The ANOVA results, organized in Table 3, showed how NDVI raw map $X_{raw}^{II}$, derived from the satellite imagery, has no accordance with the map generated from in-field measurement $V_{field}^{II}$. The difference between the means of the pixel groups (Figure 4), obtained by clustering NDVI map $X_{raw}^{II}$ by using the spatial information provided by in-field survey $V_{field}^{II}$, was found not to be significant, with obtained p-values ranging from 0.04 to 0.26 for all three considered parcels A, B, and C (Table 3). This confirms the limitations of $X_{raw}^{II}$ in directly providing reliable information regarding the status of the vineyards in this scenario, where the radiometric information reflected from the crop field could be affected by other sources (e.g., inter-row paths) that, in the case of crops grown by rows, could be predominant and could negatively affect the overall NDVI assessment. On the contrary, by using the same assessment approach, the effectiveness of the NDVI map derived from UAV imagery $Y_{UAV}^{II}$ proved to be statistically significant, with different group means in all the considered parcels and showing a favourable coherence with in-field ground truth $V_{field}^{II}$. This preliminary analysis was propedeutic to the quality assessment of the proposed new framework to refine the satellite-driven NDVI map with the RarefyNet model. The ANOVA results demonstrated how refined NDVI map $\hat{X}^{II}$ correlates with reference $V_{field}^{II}$, with small p-values ranging from 0.0015 to $3.17 \times 10^{-8}$ (Table 3), drastically improving the performance of raw satellite-driven dataset $X_{raw}^{II}$. The results presented so far prove that the proposed RarefyNet is capable of refining the raw Sentinel-2 driven map $\hat{X}^{II}$ of time period II by extracting the features from UAV-driven map $Y_{UAV}^{I}$.

**Table 3.** ANOVA results for the June (time *II*) datasets $X_{raw}^{II}$, $\hat{X}^{II}$, and $Y_{UAV}^{II}$ grouped according to ground truth vigor map $V_{field}^{II}$: raw Sentinel-2 $X_{raw}^{II}$ does not show significant differences among the vigor group means defined by the field expert with in-field measurement $V_{field}^{II}$, whilst enhanced UAV map $Y_{UAV}^{II}$ and the refined version of Sentinel-2 map $\hat{X}^{II}$ show significant differences among the group means.

| Datasets (Grouped by) | Parcel | Source | DF [1] | SS [1] | MS [1] | F-Value | *p*-Value |
|---|---|---|---|---|---|---|---|
| $X_{raw}^{II}(V_{field}^{II})$ | Parcel-A | Classes | 2 | 0.3084 | 0.1541 | 3.4582 | 0.044081 |
| | | Error | 31 | 1.3821 | 0.0445 | | |
| | | Total | 33 | 1.6905 | | | |
| | Parcel-B | Classes | 2 | 0.3938 | 0.1969 | 4.8928 | 0.010587 |
| | | Error | 63 | 2.5353 | 0.0402 | | |
| | | Total | 65 | 2.9291 | | | |
| | Parcel-C | Classes | 2 | 0.1985 | 0.0992 | 1.4555 | 0.264401 |
| | | Error | 15 | 1.0228 | 0.0681 | | |
| | | Total | 17 | 1.2213 | | | |
| $\hat{X}^{II}(V_{field}^{II})$ | Parcel-A | Classes | 2 | 0.4749 | 0.2374 | 8.0112 | 0.001568 |
| | | Error | 31 | 0.9189 | 0.0296 | | |
| | | Total | 33 | 1.3938 | | | |
| | Parcel-B | Classes | 2 | 1.3735 | 0.6867 | 22.9984 | $3.17 \times 10^{-8}$ |
| | | Error | 63 | 1.8812 | 0.0298 | | |
| | | Total | 65 | 3.2547 | | | |
| | Parcel-C | Classes | 2 | 0.7071 | 0.3535 | 11.7444 | 0.000852 |
| | | Error | 15 | 0.4515 | 0.0301 | | |
| | | Total | 17 | 1.1586 | | | |
| $Y_{UAV}^{II}(V_{field}^{II})$ | Parcel-A | Classes | 2 | 1.3608 | 0.6804 | 30.0925 | $5.46 \times 10^{-8}$ |
| | | Error | 31 | 0.7009 | 0.0226 | | |
| | | Total | 33 | 2.0617 | | | |
| | Parcel-B | Classes | 2 | 2.7135 | 1.3567 | 71.1664 | $6.87 \times 10^{-17}$ |
| | | Error | 63 | 1.2010 | 0.0190 | | |
| | | Total | 65 | 3.9145 | | | |
| | Parcel-C | Classes | 2 | 0.9447 | 0.4723 | 8.7803 | 0.002988 |
| | | Error | 15 | 0.8069 | 0.0537 | | |
| | | Total | 17 | 1.7516 | | | |

[1] DF: degree of freedom, SS: sum of squares, MS: mean square

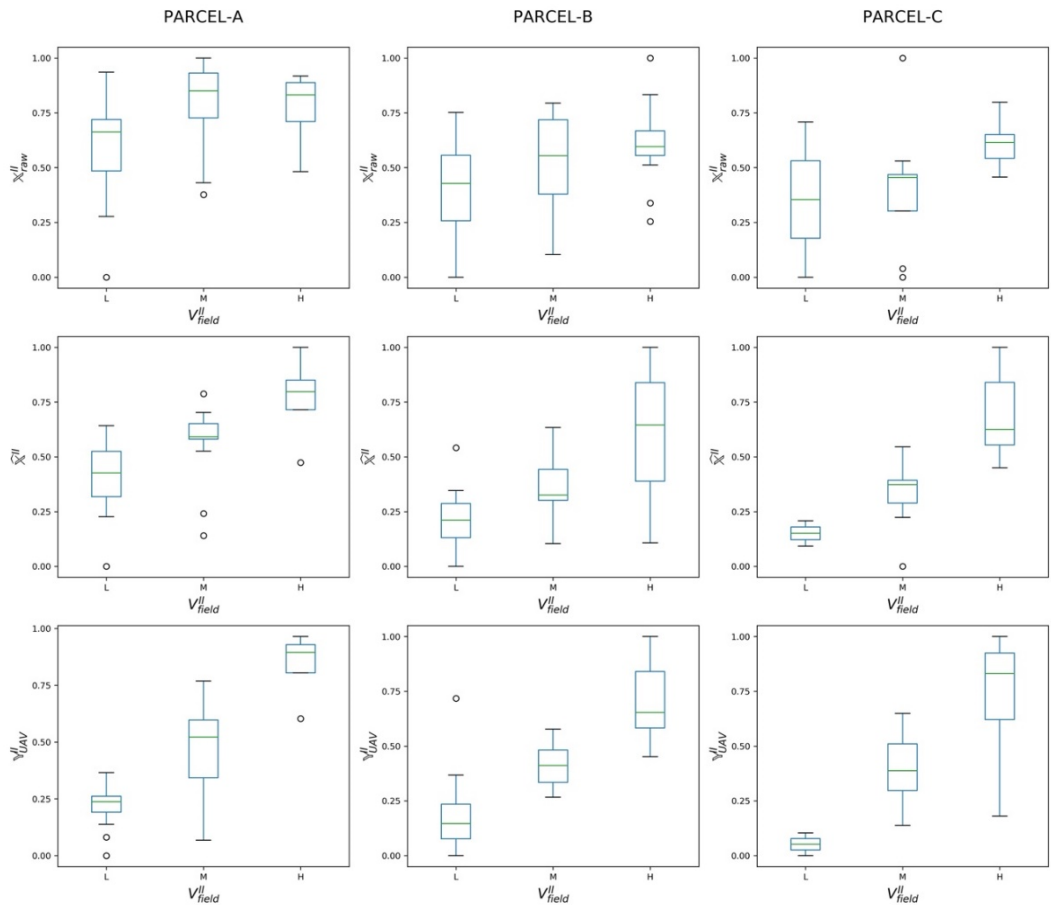

**Figure 4.** Pixel groups boxplots from raw satellite-driven map $X_{raw}^{II}$, refined satellite-driven map $\hat{X}^{II}$, and UAV-driven map $Y_{UAV}^{II}$, clustered according to the three vigor classes "L", "M", and "H" defined in map $V_{field}^{II}$. The boxplots are individually computed for each parcel (**A**, **B**, and **C**).

To extend the performed analysis to other time datasets, all the maps produced from the UAV imagery ($Y_{UAV}^{I}, Y_{UAV}^{II}, Y_{UAV}^{III}, Y_{UAV}^{IV}$) were clustered into three vigor classes by using a K-means algorithm, obtaining a set of clustered maps $Y_{UAV}^{I}, Y_{UAV}^{II}, Y_{UAV}^{III}$, and $Y_{UAV}^{IV}$. The soundness of the proposed clustering approach was demonstrated by comparing, parcel by parcel, map $Y_{UAV}^{II}$ to in-field vigor map $V_{field}^{II}$ by evaluating the Pearson correlation coefficients (Figure 3). The obtained positive values, ranging from 0.68 to 0.84, showed that the produced clustered map $Y_{UAV}^{II}$ is well correlated with $V_{field}^{II}$. This result, together with the extremely favourable ANOVA results of $Y_{UAV}^{II}$ in Table 3, makes it possible to consider the UAV-driven dataset as a robust and reliable reference in the following analysis.

The performance of the proposed RarefyNet in extending the refinement task also to other imagery from a time series, even if trained only with one single UAV-driven dataset, was thus further assessed by refining other temporal raw Sentinel-2 maps. The effectiveness of refined maps $\hat{X}^{III}$ and $\hat{X}^{IV}$ (obtained by refining maps $X_{raw}^{III}$ and $X_{raw}^{IV}$) in describing the vigor level of the vineyard expressed in reference UAV-driven maps $Y_{UAV}^{III}$ and $Y_{UAV}^{IV}$ was investigated with ANOVA. The results of this analysis, together with the ones performed on $\hat{X}^{I}$ and $\hat{X}^{II}$ for completeness, are organised in Table 4. The boxplots of the groups of pixels from the refined satellite maps ($\hat{X}^{I}, \hat{X}^{II}, \hat{X}^{III}$ and $\hat{X}^{IV}$), clustered according to the three vigor classes "L", "M", and "H" defined in the UAV-driven clustered maps $Y_{UAV}^{I}, Y_{UAV}^{II}, Y_{UAV}^{III}$ and $Y_{UAV}^{IV}$ respectively, are shown in Figure 5. The ANOVA results reported in Table 4 confirmed the good coherence of all four refined Sentinel-2 maps with their respective reference maps, with p-values showing the significance of the differences among group means. The results achieved by the performed analysis provide an opportunity to use the freely, frequently available, low resolution satellite imagery to describe the variability of vineyards by refining the satellite driven vegetation index. Refinement is done by adopting a proposed machine learning framework, which is trained with the valuable information extracted from high resolution UAV imagery and the spatial information of the satellite neighborhood pixels.

**Table 4.** ANOVA results of refined datasets $\hat{X}^I, \hat{X}^{II}, \hat{X}^{III}$, and $\hat{X}^{IV}$, grouped according to reference UAV-drive vigor maps $Y_{UAV}^I$, $Y_{UAV}^{II}$, $Y_{UAV}^{III}$, and $Y_{UAV}^{IV}$.

| Datasets (Grouped by) | Parcel | Source | DF [1] | SS [1] | MS [1] | F-Value | *p*-Value |
|---|---|---|---|---|---|---|---|
| $\hat{X}^I(Y_{UAV}^I)$ | Parcel-A | Classes | 2 | 0.7907 | 0.7907 | 32.4702 | 2.60 × 10$^{-6}$ |
| | | Error | 31 | 0.7792 | 0.0243 | | |
| | | Total | 33 | 1.5699 | | | |
| | Parcel-B | Classes | 2 | 1.39025 | 1.3902 | 78.7860 | 9.31 × 10$^{-13}$ |
| | | Error | 63 | 1.1293 | 0.0176 | | |
| | | Total | 65 | 2.5196 | | | |
| | Parcel-C | Classes | 2 | 1.1914 | 1.1917 | 113.4301 | 1.14 × 10$^{-8}$ |
| | | Error | 15 | 0.1681 | 0.0105 | | |
| | | Total | 17 | 1.3596 | | | |
| $\hat{X}^{II}(Y_{UAV}^{II})$ | Parcel-A | Classes | 2 | 0.6968 | 0.6968 | 31.9907 | 2.94 × 10$^{-6}$ |
| | | Error | 31 | 0.6970 | 0.0218 | | |
| | | Total | 33 | 1.3939 | | | |
| | Parcel-B | Classes | 2 | 1.5536 | 1.5536 | 58.4472 | 1.36 × 10$^{-10}$ |
| | | Error | 63 | 1.7012 | 0.0266 | | |
| | | Total | 65 | 3.2548 | | | |
| | Parcel-C | Classes | 2 | 0.7978 | 0.7978 | 35.3635 | 2.05 × 10$^{-5}$ |
| | | Error | 15 | 0.3609 | 0.0225 | | |
| | | Total | 17 | 1.1587 | | | |
| $\hat{X}^{III}(Y_{UAV}^{III})$ | Parcel-A | Classes | 2 | 0.4195 | 0.4194 | 13.4022 | 0.000898 |
| | | Error | 31 | 1.0015 | 0.0313 | | |
| | | Total | 33 | 1.4210 | | | |
| | Parcel-B | Classes | 2 | 0.6561 | 0.6560 | 29.8767 | 8.10 × 10$^{-7}$ |
| | | Error | 63 | 1.4054 | 0.0219 | | |
| | | Total | 65 | 2.0614 | | | |
| | Parcel-C | Classes | 2 | 0.1808 | 0.1808 | 2.1895 | 0.158372 |
| | | Error | 15 | 1.3218 | 0.0826 | | |
| | | Total | 17 | 1.5026 | | | |
| $\hat{X}^{IV}(Y_{UAV}^{IV})$ | Parcel-A | Classes | 2 | 0.2441 | 0.2441 | 4.6372 | 0.038924 |
| | | Error | 31 | 1.6846 | 0.0526 | | |
| | | Total | 33 | 1.9287 | | | |
| | Parcel-B | Classes | 2 | 0.6649 | 0.6649 | 20.8288 | 2.33 × 10$^{-5}$ |
| | | Error | 63 | 2.0431 | 0.0319 | | |
| | | Total | 65 | 2.7081 | | | |
| | Parcel-C | Classes | 2 | 0.8174 | 0.8173 | 25.5642 | 0.000117 |
| | | Error | 15 | 0.5116 | 0.0319 | | |
| | | Total | 17 | 1.3289 | | | |

[1] DF: degree of freedom, SS: sum of squares, MS: mean square

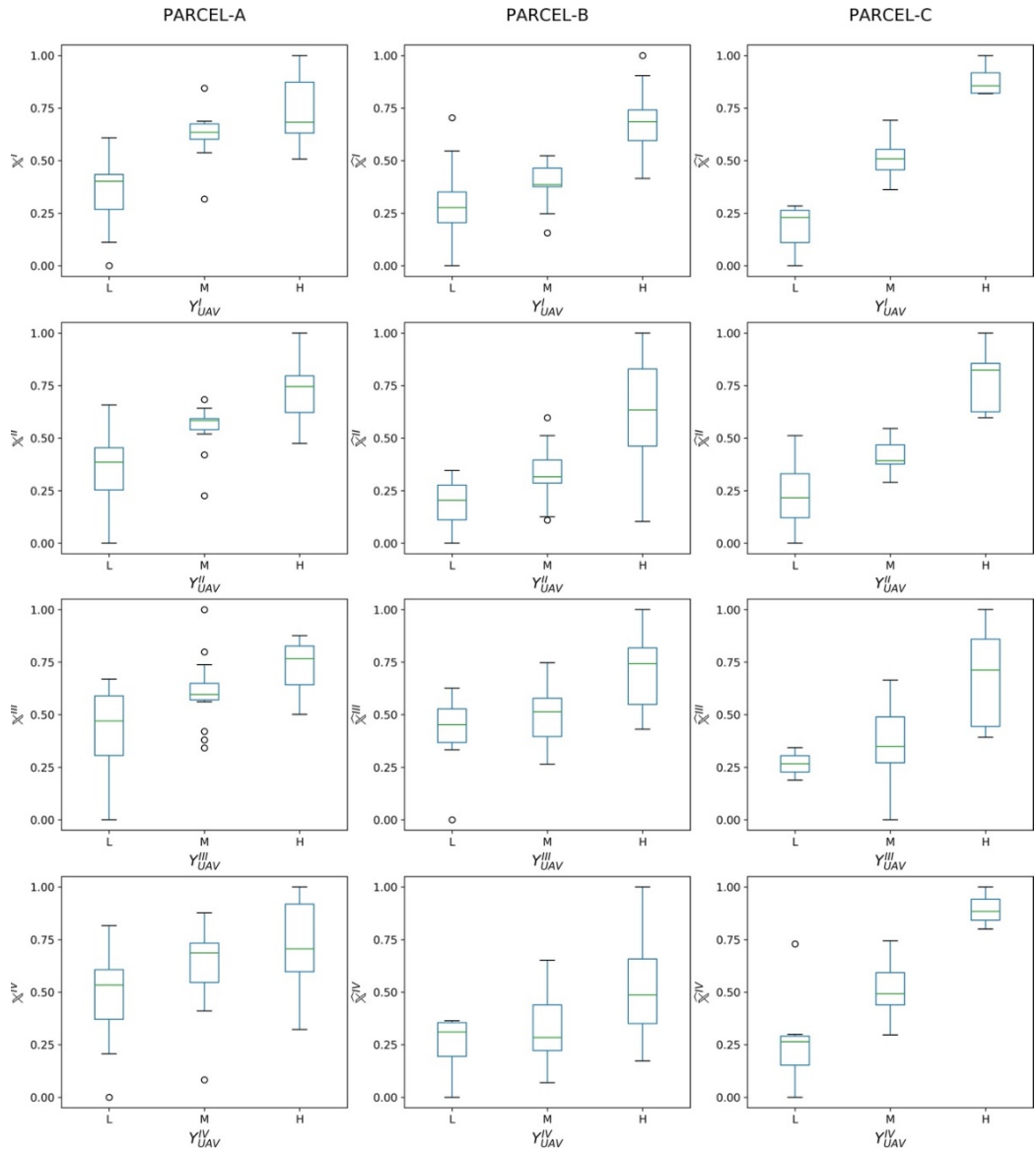

**Figure 5.** Pixel groups boxplots from refined satellite maps ($\hat{X}^I$, $\hat{X}^{II}$, $\hat{X}^{III}$, and $\hat{X}^{IV}$), clustered according to the three vigor classes "L", "M", and "H" defined in the UAV-driven clustered maps $Y^I_{UAV}$, $Y^{II}_{UAV}$, $Y^{III}_{UAV}$, and $Y^{IV}_{UAV}$, respectively. The boxplots are computed individually for each parcel (**A**, **B**, and **C**).

## 4. Conclusions

Freely available satellite imagery with low or moderate resolutions shows some limitations in specific agricultural applications, e.g., where crops are grown by rows causing biased radiometric reflectance that does not reliably describe the vegetative status. The proposed novel satellite imagery refinement framework, based on deep learning techniques, exploits information properly derived from high resolution images acquired by UAV airborne multispectral sensors. To train the convolutional neural network, only a single UAV-driven dataset is required, making the proposed approach simple and cost-effective. A vineyard in Serralunga d'Alba (Northern Italy) was chosen as a case study for validation purposes. Refined satellite-driven NDVI maps, acquired in four different periods during the vine growing season, were shown to better describe crop status with respect to raw datasets by correlation analysis and ANOVA. In addition, using a K-means based classifier, three-level vineyard vigor maps were profitably derived from the NDVI maps, which are a valuable tool for growers.

**Authors Contribution:** Conceptualization, M.C. and P.G.; methodology, V.M., L.C., and A.K.; software, V.M. and A.K.; validation, V.M. and A.K.; data curation, V.M. and A.K.; writing—original draft preparation, V.M., L.C., and A.K.; writing—review and editing, P.G. and M.C.; project administration, P.G. and M.C.; funding acquisition, P.G. and M.C. All authors have read and agreed to the published version of the manuscript.

**Funding:** This research was partially funded by the project "New technical and operative solutions for the use of drones in Agriculture 4.0" (PRIN 2017, Prot. 2017S559BB).

**Acknowledgments:** The authors would like to acknowledge Germano Ettore, owner of the winery, for hosting the experimental campaign and Iway S.r.l. for conducting the UAV flights for multispectral imaging. This work was developed with the contribution of the Politecnico di Torino Interdepartmental Centre for Service Robotics PIC4SeR (https://pic4ser.polito.it) and SmartData@Polito (https://smartdata.polito.it).

**Conflicts of Interest:** The authors declare no conflict of interest.